\documentstyle[11pt,newpasp,twoside,epsf]{article}
\def\arg#1{{\it#1\/}}

\def\hi{H{\sc i}~}
\def\halpha{\hbox{H{$\alpha$}}~}
\def\rosat{\hbox{\it ROSAT~}}
\def\chandra{\hbox{\it CHANDRA~}}
\def\xmm{\hbox{\it XMM-Newton~}}
\def \fluxunit {$\rm erg\,cm^{-2}\,s^{-1}$~}
\def \lumunit {$\rm erg\,s^{-1}$~}
\def \pcmsq {$\rm cm^{-2}$~}

\def\edcomment#1{\iffalse\marginpar{\raggedright\sl#1\/}\else\relax\fi}
\reversemarginpar

\begin{document}
\title{In search of the oldest star forming regions in Holmberg II}
 \author{J\"urgen Kerp}
\affil{Radioastronomisches Institut der Universit\"at Bonn, Auf dem H\"ugel 71, D-53121 Bonn, Germany}
\author{Fabian Walter}
\affil{California Institute of Technology, Astronomy Department 105-24, Pasadena, CA 91125, U.S.A.}

\begin{abstract}
We present an X-ray study of the irregular dwarf galaxy Holmberg II
based on deep {\it ROSAT\/} PSPC observations.  Holmberg II is one of
the most famous examples of an irregular dwarf galaxy with a disrupted
interstellar medium (ISM): expanding \hi holes are present across the
entire face of the galaxy. Huge \hi cavities of kpc extent are found
even outside current starforming regions.  We search for faint X-ray
sources (stellar end points as well as hot gas), touching the limits
of the \rosat PSPC detector and link the newly detected X-ray sources
to features detected in other wavelengths. Using an X-ray hardness
ratio color-color diagram we show that it is possible to differentiate
between thermal plasma and power-law X-ray spectra, which helps to
track down the physical nature of the individual sources.

At the X-ray flux limit of the analyzed \rosat PSPC data, the giant
\hi holes appear as devoid of coronal gas.  This can be used as a
measure for the minimum age of the \hi holes to about $t~\geq~10^7$ --
$10^8$ years.  Far outside the stellar body and positionally
coincident with some \hi holes we find faint, point like X-ray sources
which may be associated with low mass X-ray binaries (LMXBs).  This
old stellar population is undetectable in the optical regime because
they are too faint. However, in the X-ray regime they are prominent
X-ray sources, well known from Population I objects located within the
bulge of the Milky Way and globular clusters.  The search for LMXBs
might be the key for identifying the earliest stellar population in
distant galaxies.  In case of Holmberg II their presence indicates that
star formation took place across the entire gaseous disk in the past.
\end{abstract}

\section{Introduction}
Holmberg II (hereafter referred to as Ho\,II) is one of the most
famous examples of a dwarf galaxy with a disrupted neutral ISM as
traced by observations of neutral hydrogen (HI, Puche et al. 1992).
The distribution its ISM is characterized by \hi holes which are
present across the entire face of the galaxy. Huge \hi cavities of kpc
extent are found even outside Ho\,II's stellar population. Puche et
al. (1992) compiled a catalog of individual HI holes: their diameters
range between 100 -- 2000 pc, the radial velocities indicate that the
holes are still expanding. This provides a rough estimate of the ages
of the \hi holes of about $t \sim 10^7$ to $10^8$ years.  Since the
work done by Puche et al. (1992), Ho\,II has been studied in many
wavelength ranges by a number of authors (radio continuum: Tongue \&
Westpfahl, 1995; optical: Rhode et al., 1999; FUV: Stewart et al.,
2000).

The presence of the \hi cavities is usually linked to the stellar
activity within their centers (the ``standard picture''): Strong winds
of young O-- and B--stars create individual bubbles of a few pc size
while subsequent supernova events can create cavities of few tens to
hundreds of pc extent (e.g., Weaver et al. 1977).  \hi observations
indicate that in dwarf galaxies the expanding shells and holes can
grow older and to much larger sizes compared to massive spiral
galaxies.  This finding can be attributed to the shallow gravitational
potential and the absence of differential rotation in dwarf galaxies
(Walter \& Brinks 1999).

X-ray observations may serve as a new tool to check the standard
picture: the hot--gas interior of the \hi is traced by soft X-emission
which is emitted by the plasma (see Sec.~1.1).  Harder X-ray emission
originating from point sources may indicate old stellar endpoints such
as supernova remnants, low and high mass X-ray binaries and pulsars on
timescales $\>10^8$\,yrs (this will be discussed in Sec.~1.2). X--ray
observations therefore have the power to trace the global
star--formation history way further back than it is possible with,
e.g., H$\alpha$ or FUV observations (Stewart et al. 2000).

\subsection{Diffuse X-ray emission}
The cooling time of the plasma can be estimated by $t_{\rm cool}
\simeq 6340~\frac{\sqrt{T}}{n_{\rm e}}$ to be some $5\cdot 10^8$ yrs, 
assuming a temperature of $T\simeq 10^{6.4}$ K and a typical electron
density of about $n_{\rm e} = 2\cdot 10^{-2}~{\rm cm^{-3}}$ (see e.g.
Walter et al. 1998 and references therein).  In the cases of the
largest holes (with radii larger than the one--$\sigma$--scale--height
of the \hi disk -- 300\,pc in the case of Ho\,II) the expanding bubble
will break out and the hot gas interior will be vent out to the
halo. In the case of the smaller, still confined holes, the interior
X--ray emission should be detectable if the \hi column density of the
approaching \hi shell does not succeed values of $\sim
5\cdot 10^{20}$\,cm$^{-2}$ (when the soft X--emission of the hot gas
is intrinsically absorbed by the shell).

\subsection{Individual objects}
Pulsars, cooling neutron stars as well as X-ray binaries are
stellar end-points and may be observable within the interior of the
holes.  Their X-ray luminosities range between $L_{\rm X}({\rm 0.1 -
2.4 keV}) \simeq 10^{36}$--$10^{37} {\rm erg~s^{-1}}$, corresponding
to an X-ray flux (adopting a distance to Ho\,II of 3.05\,Mpc, Hoessel
et al. 1998) of $F_X \simeq 10^{-14}$--$10^{-13} {\rm
erg~s^{-1}~cm^{-2}}$, which is detectable with the \rosat PSPC
detector in deep pointed observations like the one discussed here.

In this paper we present our analysis of three pointed \rosat PSPC
observations of Ho II. In Sec.\ 2 we present the data reduction and
analysis. Our results are compiled in Sec.\ 3.

\section{\rosat observations and data analysis}
We extracted 3 pointed PSPC observations towards Ho\,II from the
\rosat archive in Munich. The \rosat PSPC data were analyzed by us 
using the EXSAS software package (Zimmermann et al. 1998).  All three
observations were merged into a single photon event file.  The total
integration time of the merged \rosat PSPC data is 22566 sec, making
Ho\,II one of deepest studied dwarf galaxies by \rosat.  The photon
events were binned into the standard \rosat \onequarter\,keV (\rosat
C-band), \threequarters\,keV (M-band), 1.5 keV (J-band) energy bands
and a total energy band, all were corrected for vignetting.  Within
the area of an individual X-ray source towards HoII (3.5 $\times$ FWHM
PSF), the absorbed X-ray flux of the extragalactic X-ray background (XRB) is
about $F_{\rm XRB} = 4 \cdot 10^{-15}$\fluxunit. To evaluate this flux
limit we assumed that the extragalactic XRB is only attenuated by the
ISM of the Milky Way.

Unfortunately, there is no straight forward way to differentiate
between an X-ray source belonging to HoII or an accidental positional
coincidence with an XRB source. However, we can reduce this ambiguity
by studying a particular X-ray source and searching for counterparts
in other wavelengths using supplementary data (which is what we will
do in the following, the complete analysis will be presented
elsewhere, Kerp et al., 2001, in prep.).

\subsection{Detection limits}
The absorbed 3-$\sigma$ X-ray flux level across the entire \rosat energy band
is $F_{\rm 3-\sigma}{\rm
(0.1\,-\,2.1\,keV)}\,=\,(4.1\,\pm\,1.4)\cdot 10^{-15}$\fluxunit. Assuming
a distance to Ho\,II of 3.05 Mpc and a galactic X-ray attenuating
column density of about $N_{\rm \hi}\,=\,3\cdot 10^{20}$\pcmsq (Hartmann
\& Burton 1997), we derive a detection luminosity threshold of $L_{\rm
3-\sigma}\,=\,(9.8\,\pm\,3.3)\cdot 10^{36}$ \lumunit.

Only very young supernova remnants or X-ray binaries are typical
sources with luminosities exceeding this threshold. The presented
\rosat PSPC X-ray data therefore traces only two extreme populations
of the star formation in HoII: the actual ($t < 10^5$ yr, i.e. young
supernovae) or the very old population ($t > 10^7$ yr, pulsars,
X-ray binaries). The intermediate star population can be
traced via \halpha and far-ultraviolet (FUV) observations (Stewart et
al. 2000). Combining all data provides a unique insight on the star
formation history of Ho\,II during the last 100 Myrs.

\begin{figure}
\plotfiddle{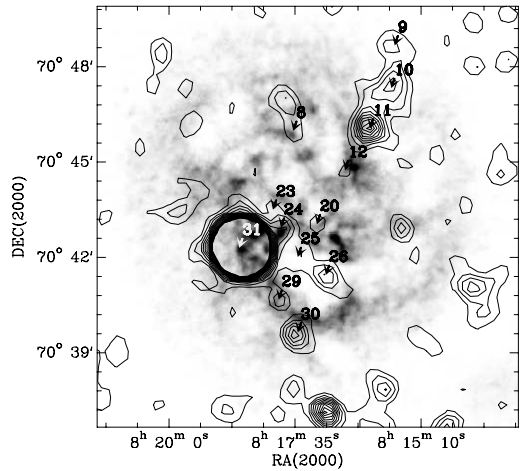}{6cm}{270}{40}{40}{-100}{210}
\caption{\rosat PSPC total (0.1 keV -- 2.1 keV) X-ray intensity
distribution superimposed as contours on the \hi 21-cm map of HoII.
The contour lines start at the 3-$\sigma$ level in steps of
2-$\sigma$.  The arrows point to the center of each X-ray source.}
\end{figure}

\section{Results}

In total we detected 31 significant X-ray sources within the extent of
the \hi distribution of HoII. To avoid any confusion with bright
extragalactic X-ray background sources, we decided to further study
only those 13 X-ray sources which are identified in at least one
additional frequency range.  The supplementary data give us also a
better handle to constrain the nature of the X-ray sources in question.
We tried to constrain the X-ray emission process of the 13 sources by
studying the spectral properties.  Because most of the X-ray sources
are rather faint, it is impossible to extract a significant X-ray
spectrum from the \rosat data.  To constrain the X-ray spectrum we
produced a multi source X-ray hardness ratio color-color diagram,
which is described in detail below.

\subsection{The X-ray color--color diagram}
To identify the emission mechanism of an X-ray source the standard
procedure is to extract a spectrum from the \rosat PSPC data.
%The brightest source associated with
%HoII (here labeled as \#31) has been studied by Zezas et al.
%(\cite{zezas}). Zesas et al. presented an analysis of this bright source;
%however, due to the large number of relevant parameters, even the analysis of 
%this high signal--to--noise X-ray spectrum is not without its ambiguities.
In case of faint X-ray sources ($F_{\rm X} \simeq 10^{-15}$\fluxunit)
-- which are discussed here -- the insufficient signal--to--noise
ratio does not allow to apply this standard procedure.  However, the
count rates within the broad \rosat energy bands contain significant
information on the X-ray source spectrum. We therefore calculate the
X-ray hardness ratios to constrain the emission process.  In Fig.~2 we
plotted the \rosat hardness-ratio 1 (HR1) ($\frac{{\rm M+J-C}}{{\rm
C+M+J}}$) versus the \rosat hardness-ratio 2 (HR2) ($\frac{{\rm
J-M}}{{\rm M+J}}$).  The hardness ratios depend strongly on the
intrinsic source spectrum but also on the amount of X-ray attenuating
matter along the line of sight.  The stronger the photoelectric
absorption the harder the resulting X-ray spectrum. Each individual
X-ray spectrum has its own trajectory in this X-ray hardness ratio
color-color diagram. This is similar to the change of the stellar color -- reddening -- of an optical
color-color diagram by dust attenuation. In Fig.~2 we plotted
trajectories for power-law (solid lines) and thermal plasma spectra
(dashed lines) with increasing column density as parameter (from left
to right $N_{\rm \hi} = 0\cdot 10^{20}~{\rm cm^{-2}}$, $2\cdot
10^{20}~{\rm cm^{-2}}$, $3\cdot 10^{20}~{\rm cm^{-2}}$ and $4\cdot
10^{20}~{\rm cm^{-2}}$).  The most upper and left trajectory of each
hardness ratio trajectory represents the the unabsorbed situation. The
labeled trajectories (line with $N_{\rm \hi} = 3\cdot 10^{20}{\rm
cm^{-2}}$, representing the absorbing foreground column density
belonging to the Milky Way) give the spectral slope in case of the
power-law spectra (solid lines) and the log($T$[K]) in case of the
thermal plasma spectra. Therefore a soft X-ray spectrum source can be
found in the lower left part of the X-ray hardness ratio color-color
diagram, whereas highly absorbed or hard X-ray sources populate the
upper right part of the diagram. The harder the X-ray spectrum, the
larger $HR1$ and $HR2$. All these trajectories, with $N_{\rm \hi}$ as
parameter, are based on the assumption that the X-ray sources of
interest are at least attenuated by the interstellar medium belonging
to the Milky Way ($N_{\rm
\hi}{(\rm Milky Way)} = 3\cdot 10^{20}~{\rm cm^{-2}}$).

\begin{figure}
\plotfiddle{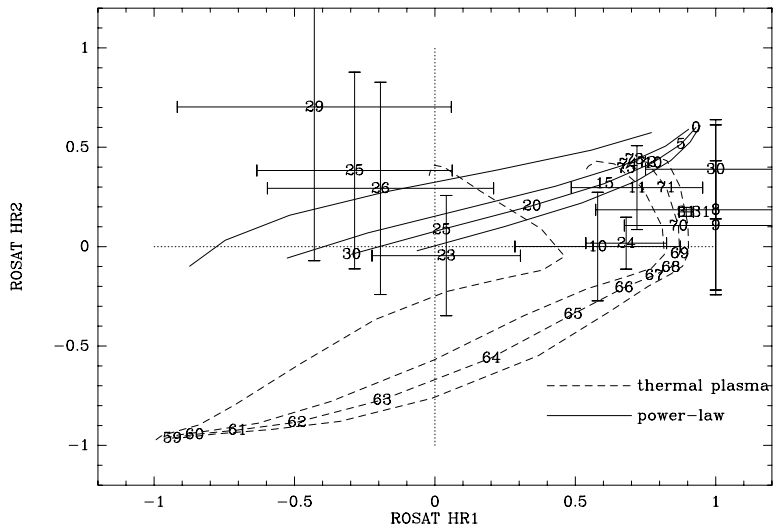}{6cm}{270}{40}{40}{-180}{210}
\caption{Hardness ratio color-color diagram of the X-ray sources
brighter than $F_{\rm X} \geq 10\cdot 10^{-15}$\fluxunit.
The X-ray sources are labeled according to the numbers given in Fig. 1.
The solid lines represent the hardness ratio trajectories
for power-law X-ray spectra, while the dashed lines mark the thermal plasma
spectra.
The trajectories have as parameter the X-ray absorbing column density.
Starting most left with the un-absorbed situation ($N_{\rm \hi} = 0\cdot
10^{20}{\rm cm^{-2}}$, with steps $2\cdot10^{20}{\rm cm^{-2}}$,
$3\cdot10^{20}{\rm cm^{-2}}$, $4\cdot10^{20}{\rm cm^{-2}}$).
The numbers along the $N_{\rm \hi} = 3\cdot10^{20}{\rm cm^{-2}}$
trajectory give the energy index in case of the power-laws and the log($T$[K]) n case of the thermal plasma spectra.}
\end{figure}

\subsection{X-ray sources with optical or radio counterparts}

The X-ray hardness ratio color-color diagram offers a very sensitive
tool to get information on the X-ray emitting process even of faint
sources. The faintest source analyzed has an X-ray flux of only
$F_{\rm X}{\rm (0.1-2.4\,keV)}\,=\,7.3\cdot10^{-15}$ \fluxunit.

However, the X-ray data alone do not allow a unique determination of
the nature of the sources. Combining all available information from
the X-ray to the low-frequency radio regime, we tried to constrain the
emission process of each particular X-ray source.

Five X-ray sources are located within the interior of the \hi holes
cataloged by Puche et al. (1992).  Two of them appear to be
associated with supernova remnants (\#23 and \#31), one can be identified
as young star-forming regions (\#20). The remaining two are most likely
stellar end-points, in particular LMXBs (\#25 and
\#26).

Four of the selected X-ray sources are located towards high column density
regions of HoII. One may be associated with a stellar cluster hosting
developed stars with bright X-ray emitting coronae (\#8), two with young
star forming regions (\#12 and \#24) and one with an X-ray
 binary system (\#29).

Finally, we found 4 X-ray sources which are located far outside the
stellar body and close to the rims of the \hi column density
distribution of HoII. All these sources represent most likely stellar
end-points, either supernova remnants or X-ray binaries.

Some of the \halpha regions of HoII are associated with X-ray
emission, however there is no strong correlation. We should keep in
mind, however, that the X-ray detection threshold is very high, which
leads to selecting only the brightest star forming regions. The much
more sensitive \chandra and \xmm observation will overcome this
instrumental limitation.

LMXBs (tracers of the past star formation) are also found well outside
the stellar body of HoII (\#25, \#26 and \#29). This indicates past star
formation at large galactocentric radii. The giant \hi holes may
therefore be indeed residuals from energetic events associated with
the stellar activity of HoII in the past (on timescale $t > 10^9$
yrs).  This would imply that LMXB are good markers for sites of old
star formation. Because they are faint in the optical they are hardly
detectable even with the {\it Hubble Space Telescope\/} at Mpc
distances.  In the X-ray domain however, they are prominent and bright
X-ray sources with a hard spectrum.  The \rosat PSPC data presented
here is unfortunately not sensitive enough to compile a complete
inventory of LMXBs in HoII.  However with \chandra and \xmm LMXBs will
be easy to detect (even in case they are deeply embedded LMXBs within
the ISM of the host galaxy).  This opens the exciting possibility to
search for LMXBs within the todays ``empty'' \hi holes of HoII and
other dwarf galaxies. In fact, only a few tens of ksec integration
time with \xmm is needed to reach the necessary flux level.  The most
critical step is to separate their X-ray emission from that of XRB
sources which is unrelated to the galaxy of interest. The improved
spectral resolution of the CCD--detectors on-board of \chandra and
\xmm will provide additional information to improve the presented
X-ray hardness color--color diagram to identify the nature of the
X-ray sources.

\end{document}